\documentclass[prd,aps,prl,twocolumn,nofootinbib,superscriptaddress,reprint]{revtex4-1}
\usepackage{amsmath,amssymb,amsthm,bm,braket,ascmac,bbm}
\usepackage{graphicx}
\usepackage{color} 
\usepackage{hyperref}
\usepackage[normalem]{ulem}

\theoremstyle{definition}

\begin{document}

\title{A Natural Model of Spontaneous CP Violation
}
\author{Sudhakantha Girmohanta}
\affiliation{C. N. Yang Institute for Theoretical Physics and Department of Physics and Astronomy,
Stony Brook University, Stony Brook, New York 11794, USA}

\author{Seung J. Lee}
\affiliation{Department of Physics, Korea University, Seoul, 136-713, Korea}
\affiliation{CERN, Theoretical Physics Department}

\author{Yuichiro Nakai}
\affiliation{Tsung-Dao Lee Institute and School of Physics and Astronomy, Shanghai
Jiao Tong University, 800 Dongchuan Road, Shanghai 200240, China}

\author{Motoo Suzuki}
\affiliation{Tsung-Dao Lee Institute and School of Physics and Astronomy, Shanghai
Jiao Tong University, 800 Dongchuan Road, Shanghai 200240, China}

\begin{abstract}
We examine the possibility of building a natural non-supersymmetric model of spontaneous CP violation equipped with the Nelson-Barr (NB) mechanism
to address the strong CP problem.
Our approach is to utilize a doubly composite dynamics where the first confinement of the CFT occurs at  the scale of spontaneous CP violation (SCPV)
and the second confinement at the TeV scale. A holographic dual description of this 4D set-up via a warped extra dimension with three 3-branes provides an explicit realization of this idea.
In this model, radiative corrections to the strong CP phase
are well under control,
and the coincidence of mass scales, which we generally encounter in NB models, is addressed. 
Our model also provides an explanation to the quark Yukawa hierarchies,
and a solution to the gauge hierarchy problem just as in the usual
Randall-Sundrum model with the Higgs being localized on the TeV brane.
\end{abstract}

\maketitle

\section{Introduction}

The strong CP problem, namely why CP violation in the strong interaction is so small, $i.e.$ $\bar{\theta} \lesssim 10^{-10}$, is one of the outstanding puzzles in particle physics.
There is no symmetry protecting $\bar{\theta}$ to be small within the Standard Model (SM),
similar to the case of the Higgs mass.
But, even worse than the case of the Higgs mass, there is no known direct anthropic solution to it. 
The most well known solution is the Peccei-Quinn solution with the QCD axion~\cite{Peccei:1977hh, tHooft:1976rip}. As an alternative, 
spontaneous violation of the CP symmetry~\cite{Beg:1978mt, Georgi:1978xz, Mohapatra:1978fy, Segre:1979dx, Barr:1979as, Nelson:1983zb, Barr:1984qx, Barr:1984fh, Babu:1989rb, Barr:1991qx, Kuchimanchi:2010xs} is a promising paradigm of physics beyond the SM, 
which also realizes
the Cabibbo-Kobayashi-Maskawa (CKM) phase of the quark-mixing matrix.\footnote{
Spontaneous CP violation also plays a role in suppressing contributions to electric dipole moments in supersymmetry
\cite{Nir:1996am,Aloni:2021wzk,Nakai:2021mha}.
}
The most popular realization of the idea is known as the Nelson-Barr (NB) mechanism
\cite{Nelson:1983zb,Barr:1984qx,Barr:1984fh}
where the QCD vacuum angle $\bar{\theta}$ is protected under the spontaneous CP violation
at the classical level by the structure of an extended quark mass matrix.
A minimal model  was proposed by Bento, Branco and Parada (BBP)~\cite{Bento:1991ez},
which introduces a vector-like pair of $SU(2)_L$ singlet quarks
and SM singlet scalar fields breaking the CP symmetry spontaneously.

However, the NB models encounter several theoretical challenges (see, for example, a well documented summary in~\cite{Dine:2015jga}).
First, to suppress higher-dimensional operators destroying the NB mechanism,
the scale of spontaneous CP violation must be hierarchically smaller than a cutoff scale of the theory.
In non-supersymmetric models, fine-tuning caused by quadratically divergent radiative corrections
to the masses of the CP-violating scalar fields is even worse than that of the original strong CP problem.
Second, although the NB mechanism works at the classical level,
an unacceptably large value of $\bar{\theta}$ is generated radiatively in general.
These issues can be addressed when we consider supersymmetric NB models
where supersymmetry breaking is provided by gauge mediation.
Loop effects on $\bar{\theta}$ in this case have been estimated in ref.~\cite{Fujikura:2022sot}
(see also refs.~\cite{Dine:1993qm,Hiller:2001qg,Hiller:2002um,Dine:2015jga,Evans:2020vil}).
Another approach to protect a hierarchically small scale of spontaneous CP violation is
to utilize {\it a} strong dynamics which makes the CP-violating scalar fields composite, while additional gauge symmetries may be introduced to suppress the contribution to $\bar\theta$ (see $e.g.$ \cite{Vecchi:2014hpa,Valenti:2021xjp} for recent works).

In this paper, we explore a non-supersymmetric NB model based on the BBP model~\cite{Bento:1991ez}
to realize
a hierarchically small scale of spontaneous CP violation and to sufficiently suppress radiative corrections to $\bar{\theta}$.
We introduce a warped extra dimension with three 3-branes
whose typical energy scales are given by the Planck scale, the scale of spontaneous CP violation
and the TeV scale, respectively.
Warped extra dimension models with multiple branes have been explored in various studies~\cite{Lykken:1999nb,Hatanaka:1999ac,Kogan:1999wc,Oda:1999di,Oda:1999be,Dvali:2000ha,Gregory:2000jc,Kogan:2000cv,Mouslopoulos:2000er,Kogan:2000xc,Pilo:2000et,Choudhury:2000wc,Kogan:2001qx, Mouslopoulos:2001uc,Kogan:2001wp,Moreau:2004qe,Agashe:2016rle,Agashe:2016kfr,Csaki:2016kqr,Fichet:2019owx,Fuentes-Martin:2020pww,Cai:2021nmk, Lee:2021slp, Fuentes-Martin:2022xnb} and the radion stabilization for multiple branes has been established~\cite{Lee:2021wau}.
According to the AdS/CFT correspondence
\cite{Maldacena:1997re,Gubser:1998bc,Witten:1998qj},
a 5D three-brane model is dual to a nearly-conformal strongly-coupled 4D field theory
where spontaneous breaking of the conformal symmetry takes place via confinement twice~\cite{Agashe:2016rle}.
Three branes are essential to forbid dangerous operators leading to a large $\bar{\theta}$.
We find that the model can solve the strong CP problem without fine-tuning.
Furthermore, the model accommodates the mechanism to explain fermion mass hierarchies
through localized profiles of bulk SM fermions
\cite{Grossman:1999ra,Gherghetta:2000qt}.
This feature does not exist in supersymmetric models
and delivers a unique advantage to our model.

The rest of the paper is organized as follows.
In section~\ref{NB}, we review the NB mechanism and the BBP model.
Section~\ref{model} presents our warped extra dimension model with three 3-branes
to solve the strong CP problem without fine-tuning.
In section~\ref{theta_correction}, we discuss corrections to $\bar{\theta}$ in our model. 
Section~\ref{conclusion} is devoted to conclusions.
Details about a 5D fermion in the three 3-brane setup are summarized in appendix~\ref{bulkfermion}.

\section{Nelson-Barr mechanism}\label{NB}

Let us start with a review of the NB mechanism~\cite{Nelson:1983zb,Barr:1984qx,Barr:1984fh}
and the BBP model~\cite{Bento:1991ez} as its simplest realization.
We assume that our Lagrangian is invariant under the CP transformation.
The CP symmetry is only spontaneously broken by 
vacuum expectation values (VEVs) with nonzero phases of SM singlet complex scalar fields $\eta_{a} \, (a \geq 2)$.
We also introduce a vector-like pair of quarks $q_d,\bar q_d$
where $\bar q_d$ is in the same representation as the right-handed down-type quarks $\bar d$
under the SM gauge group.
To suppress unwanted terms, a $\mathbb{Z}_N$ $(N \geq 3)$ symmetry is imposed.
The newly introduced fields $q_d,\bar q_d,\eta$ have charges $+1,-1,-1$, respectively, $i.e.$ 
they transform by $q\to e^{2\pi i/N}q\ ,~\bar q\to e^{-2\pi i /N}\bar q\ ,~\eta_a\to e^{-2\pi i /N} \eta_a$,
while the SM fields are neutral.%
\footnote{We may consider a $\mathbb{Z}_2$ symmetry instead of the $\mathbb{Z}_N$ $(N \geq 3)$.
In that case, a CP phase can be provided into the SM sector with only one $\eta_a$ $(a=1)$ by $\mathcal{L}\sim \sum_{f} (a_f \eta_1+a'_f\eta_1^*)q_d \bar d_f $.
In addition, $e.g.$ the first term in Eq.~\eqref{eq:danger_2} is not dangerous anymore with only one $\eta_a$,
but now we have another dangerous operator like $\mathcal{L} \supset  \frac{\eta_1}{\Lambda}\eta_1 \bar q_d q_d$.}
The relevant part of the Lagrangian is then given by
\begin{align}
\label{eq:lagrangian}
\mathcal{L}_{\rm NB}=\mu_d q_d\bar q_d+\sum_{a,f}a_{af}^d \eta_a q_d\bar d_f+\sum_{f,f'}Y^d_{ff'} HQ_f \bar d_{f'}\ ,
\end{align}
where $\mu_d$ is a real parameter with a mass dimension one, $f=1,2,3$ denotes the flavor index
and $a_{af}^d$, $Y^d_{ff'}$ are real dimensionless couplings.
The third term is the ordinary Yukawa interaction with the left-handed quark $Q$ and the Higgs field $H$.
We focus on the parameter space with $\mu_d, a_{af}^d\langle \eta_a\rangle > 1 \,{\rm TeV}$.
Then, the first and second terms in Eq.~\eqref{eq:lagrangian} make one linear combination of $\bar q$ and $\bar d$ massive,
while the other combinations give the SM down-type quarks that obtain nonzero masses from
the third Yukawa term.

The observed CKM phase is provided when $\mu$ is comparable to $|a^d_{af}\eta_a|$~\cite{Vecchi:2014hpa,Valenti:2021rdu}.
On the other hand, the QCD vacuum angle,
\begin{align}
\bar\theta = \theta - \arg (\det \hat M_u \cdot \det \hat  M_d ) \, ,
\end{align}
is not generated at the tree level.
Here, $\hat  M_u$ is the $3 \times 3$ up-type quark mass matrix
and $\hat  M_d$ denotes the $4 \times 4$ down-type quark mass matrix
including the vector-like quark components,
\begin{align}
\label{eq:M_d}
\hat M_d=
\left(
\begin{array}{cc}
\mu_d & \sum_a a^d_{af}\langle \eta_a\rangle \\[0.5ex]
0 & Y^d_{ff'} \langle H\rangle 
\end{array}
\right) .
\end{align}
Note that only one off-diagonal block includes complex phases due to $\langle\eta_a\rangle$
while the other off-diagonal block is zero, leading to a real $\det{\hat M_d}$.

The BBP model, however, suffers from corrections to the down-type quark mass $\hat M_d$
which generally induce a non-zero $\bar \theta$.
The discrete $\mathbb{Z}_N$ symmetry cannot forbid higher dimensional operators such as
\begin{align}
\label{eq:danger_2}
\mathcal{L} \supset  \frac{\eta_b^*}{\Lambda}\eta_a \bar q_d q_d+\frac{\eta_a^*}{\Lambda} HQ \bar q_d\ ,
\end{align}
where $\Lambda$ denotes a UV cutoff scale and coefficients of the terms are omitted.
If $\Lambda$ is given by the Planck scale $M_{\rm Pl}$,
the absolute values of $\langle\eta_a\rangle$ must be much smaller than
the Planck scale,
$|\langle\eta_a\rangle|/M_{\rm Pl} < 10^{-10}$, to satisfy $\bar \theta < 10^{-10}$.
Since the theory is non-supersymmetric, this leads to another naturalness problem
which is more serious than the strong CP problem itself. 
Furthermore, a large $\bar \theta$ can be generated by radiative corrections.
We can write down
\begin{align}
\label{eq:danger_3}
\mathcal{L} \supset \gamma_{ab} \eta_a^*\eta_b H^\dagger H+\gamma_{abcd}\eta_a\eta_b\eta^*_c\eta^*_d
+ \rm h.c. \, ,
\end{align}
where $\gamma_{ab}$ and $\gamma_{abcd}$ are dimensionless coupling constants.
The first and second terms in Eq.~\eqref{eq:danger_3}, respectively, lead to corrections to $\bar \theta$
at one and two loops with the interactions in Eq.~\eqref{eq:lagrangian}, which may be estimated as~\cite{Dine:2015jga}
\begin{align}
&\delta \bar \theta\sim \left|\frac{\gamma_{bc} a^d_{af}a^d_{bf}\eta_a\eta^*_c}{16\pi^2 M_{\rm CP}^2}\right|\ ,\\[1ex]
\label{eq:delta_theta_gamma}
&\delta \bar \theta\sim \left|\frac{g^2\gamma_{abcd} a^d_{af}a^d_{cf}\eta_b^*\eta_d}{(16\pi^2)^2 M_{\rm CP}^2}\right|\ ,
\end{align}
where $M_{\rm CP}$ represents the scale of spontaneous CP violation,
$g$ denotes a SM gauge coupling
and the complex phases of $\eta_a$ are assumed to be $\mathcal{O}(1)$. 
In the next section, we will present a simple solution to suppress these dangerous contributions to $\bar \theta$.

\section{The model}\label{model}

We consider a 5D model whose extra dimension is compactified on the $S^1/Z_2$ orbifold.
The spacetime geometry is described by the metric,
\begin{align}
\label{eq:bc_metric}
ds^2=e^{-2A(y)} \eta_{\mu\nu}dx^\mu dx^\mu-dy^2 .
\end{align}
Here, $y \in [0, y_{\rm IR}]$ denotes the fifth coordinate
and $A(y)$ is a function of $y$.
Three 3-branes are introduced: two of them are located at the orbifold fixed points $y=0,y_{\rm IR}$
which we call the UV and IR branes, respectively.
The other brane is at $y=y_{\rm I} \,\, (0<y_{\rm I}<y_{\rm IR})$, which we call the intermediate brane.
The subregions $1,2$ are defined as the bulk region of $0<y<y_{\rm I}$
and the region of $y_{\rm I}<y < y_{\rm IR}$, respectively.
The warp factor $A(y)$ is different for each subregion,
\begin{align}
A (y) =
\left\{
\begin{array}{c}
k_1y~~~~~~~~~~~~~~~~~~~~~~~\text{(subregion 1)}\\[1ex]
k_2(y-y_{\rm I})+k_1y_{\rm I}~~~~~\text{(subregion 2)}
\end{array}
\right. ,
\end{align}
where $k_{1,2}$ are constant with mass dimension $1$ and we assume $k_2>k_1$
to avoid a tachyonic radion corresponding to the intermediate brane tension.
See ref.~\cite{Lee:2021wau} for a more detail discussion about the three 3-brane setup
and the radion stabilization.

To implement the NB mechanism, we assume that all the complex scalar fields $\eta_a$
are localized at the intermediate brane.
These fields obtain CP-violating VEVs around the mass scale $\sim k_1 e^{-k_1y_{\rm I}}$ owing to the warped geometry.

\begin{figure}
\centering
\hspace{1cm}
   \includegraphics[width=1\linewidth]{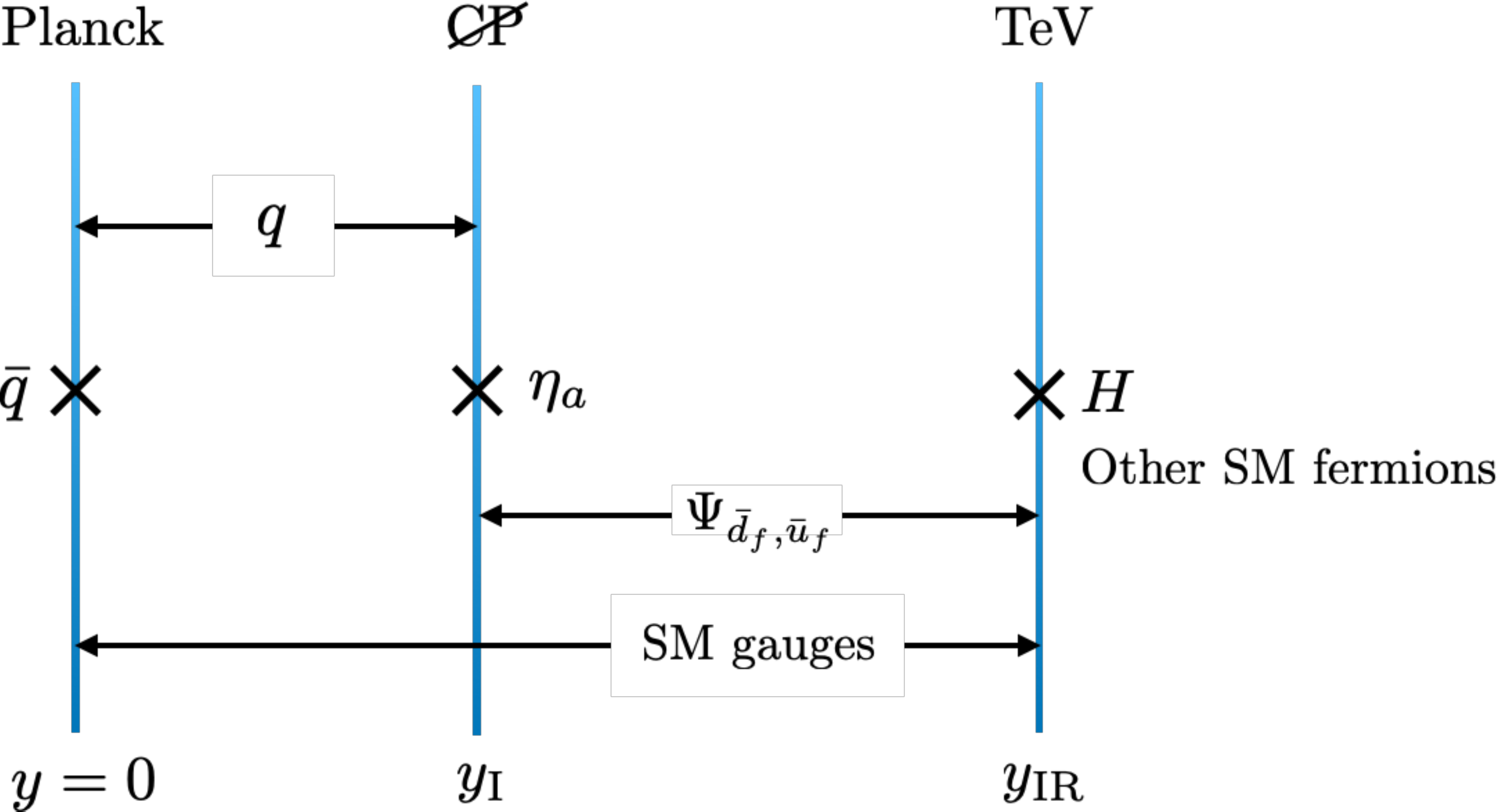}
   \vspace{0.2cm}
   \caption{A schematic picture of the setup.
   Here, $q$ denote $q_u$ and $q_d$ while $\bar q$ denote $\bar q_u$ and $\bar q_d$.}
\label{fig:setup}
\end{figure}

As for the vector-like quarks, we introduce $q_d$ as a chiral zero mode from a 5D fermion $\Psi_{q_d}$ living in the subregion $1$
and $\bar q_d$ as a 4D fermion localized at the UV brane.
The 5D down-type quarks $\Psi_{\bar d_f}$ including $\bar d_f$ as zero modes live in the subregion $2$.
In addition to those quarks, we introduce another vector-like pair of quarks
by putting a 5D fermion $\Psi_{q_u}$ in the subregion $1$ and a 4D fermion $\bar q_u$,
which has the same charge as the right-handed up-type quarks, at the UV brane.
The 5D up-type quarks  $\Psi_{\bar u_f}$ including $\bar u_f$ as zero modes live in the subregion $2$.
The SM gauge fields live in the whole bulk region.
The other SM fermions, and the SM Higgs field live on the IR brane.
A schematic picture of our model is shown in Fig.~\ref{fig:setup}.
The reason that $SU(2)_L$ doublet quarks $Q$ live only on the IR brane is 
to avoid potentially large radiative corrections to $\bar\theta$.
For the same reason, we assume that $\Psi_{\bar d_f}$ and $\Psi_{\bar u_f}$ live only in the subregion $2$ instead of the whole bulk region.
It is also noted that a naive extension of the BBP model only requires $\Psi_{\bar d_f}$ live in the subregion $1$ and
the right-handed up-type quarks $\bar u$ live at the IR brane.
However, again, to suppress corrections to $\bar\theta$,  we do not take this option.
We discuss in more detail about the motivation to adopt our setup in the next section.

The first term in Eq.~\eqref{eq:lagrangian} is introduced by a UV brane-localized coupling,
\begin{align}
\label{eq:mu_0}
S&\supset\int d^4x \sqrt{|g_{\rm in}|}\,  \hat\mu_d \psi_{q_d} \bar q_d + {\rm h.c.}\, \Bigl|_{y=0} \nonumber \\
&\supset \int d^4x\,  \mu_d q_d \bar q_d + {\rm h.c.}
\end{align}
Here, $g_{\rm in}$ denotes the determinant of the induced metric on the UV brane, $i.e.$ $|g_{\rm in}|=1$,
and $\psi_{q_d}$ is the left-handed component of $\Psi_{q_d}$, which contains $\psi_{q_d} (x,y)\ni f_{q_d}(y) q_d(x)$
with the bulk profile of the zero mode $f_{q_d}(y)$ whose normalization is determined so that
$q_d$ is canonically normalized. 
The parameter $\hat\mu_d$ is naturally given by $\hat\mu_d\sim k_1^{1/2}$,
while the parameter $\mu_d$ can be hierarchically different from $k_1$ depending on the bulk profile of $q_d$,
\begin{align}
\label{eq:mu}
\mu_d= \hat\mu_d f_{q_d}(0) = \hat\mu_d \sqrt{\frac{(1-2c_{q_d})k_1}{e^{(1-2c_{q_d})k_1y_{\rm I}}-1}}\ ,
\end{align}
with a dimensionless bulk mass parameter $c_{q_d}$ when the bulk mass of $\psi_{q_d}$ is parametrized in units of $k_1$.
See appendix~\ref{bulkfermion} for more details about a 5D fermion in the three 3-brane setup.
The second term of Eq.~\eqref{eq:lagrangian} is obtained on the intermediate brane,
\begin{align}
\label{eq:mix_qd}
S &\supset \int d^4x \sqrt{|g_{\rm in}|}\,   \sum_{a,f}\hat a_{af}^d \eta_a\psi_{q_d} \psi_{\bar d_f} + {\rm h.c.} \Bigl|_{y=y_{\rm I}} \nonumber \\
&\supset \sum_{a,f}\int d^4x\,a_{af}^d \eta_a q_d \bar d_f + {\rm h.c.}\ ,
\end{align}
where $\hat a_{af}^d$ is a constant with mass dimension $-1$ and
$\psi_{\bar d_f}$ denotes the right-handed component of the 5D fermion $\Psi_{\bar d_f}$.
The order of magnitude of $|a^d_{af}\eta_a|$ may be suppressed compared to $|\eta_a|\approx k_1e^{-k_1y_{\rm I}}$ depending on the bulk profiles of $q_d$ and ${\bar d_f}$.
As discussed in the previous section,
the observed CKM phase can be provided for $\mu_d\approx |a^d_{af}\eta_a|$.
In fact, in our setup, the mass parameter $\mu_d$ can be much smaller than the Planck scale
by assuming that $q_d$ is localized toward the intermediate brane,
so that we can realize $|a^d_{af}\eta_a|\approx \mu_d$. The Yukawa interaction terms in Eq.~\eqref{eq:lagrangian} are obtained on the IR brane
where the Higgs is localized.%
\footnote{Note that there does not exist terms involving only the down-type quarks and $\eta_a$,
\begin{align}
\label{eq:dd}
S\sim \int d^5 x \sqrt{|g_{\rm in}|} ( \eta_a \bar \Psi_{\bar d_f}\Psi_{\bar d_f} \delta(y-y_{\rm I})+ {\rm h.c.})\ ,
\end{align} 
because of the Dirichlet boundary condition for $\Psi_{\bar d_f}$ on the intermediate brane. 
 }
We also have the similar couplings for $\Psi_{\bar u_f}$, $\Psi_{q_u}$ and $\bar q_u$, $i.e.$
\begin{align}
S&\supset\int d^4x \sqrt{|g_{\rm in}|}\,  \hat\mu_u \psi_{q_u} \bar q_u + {\rm h.c.}\, \Bigl|_{y=0} \nonumber \\
&\supset \int d^4x\,  \mu_u q_u \bar q_u + {\rm h.c.}\ ,\\
S &\supset \int d^4x \sqrt{|g_{\rm in}|}\,   \sum_{a,f}\hat a_{af}^u \eta_a\psi_{q_u} \psi_{\bar u_f} + {\rm h.c.} \Bigl|_{y=y_{\rm I}} \nonumber \\
\label{eq:mix_qu}
&\supset \sum_{a,f}\int d^4x\,a^u_{af} \eta_a q_u \bar u_f + {\rm h.c.}
\end{align}
Here, $\hat\mu_u\sim k^{1/2}$, $\mu_u$ is a mass parameter determined via $\hat\mu_u$, and $a^u_{af}$ are dimensionless couplings.

We encounter the coincidence of two apparently uncorrelated mass scales,
namely $\mu$ and $a_{af}\langle\eta\rangle$,
in a generic NB model \footnote{See~\cite{Valenti:2021xjp} for a recent attempt to address the coincidence problem.}.  However, in our framework, the existence of the intermediate brane together with the localization of the $q$-field
towards that brane provides 
a prescription
to the coincidence problem.
The intermediate brane naturally provides the mass scale of $\langle\eta\rangle$.
Meanwhile, the overlap of the UV brane-localized $\bar q$
and (quasi-)intermediate brane-localized $q$ determines $\mu$, which is 
controlled by the intermediate brane mass scale as seen in Eq.\,\eqref{eq:mu} for $c_q=-\mathcal{O}(1)$. 
Furthermore, 
in terms of the dual 4D description, $\bar q$ is elementary whereas $q$ is (mostly) composite. 
Hence, the prescription to the coincidence problem is similar in nature to the solution for the hierarchy problem
in warped extra dimension models.

\begin{figure}
\centering
\hspace{1cm}
   \includegraphics[width=1\linewidth]{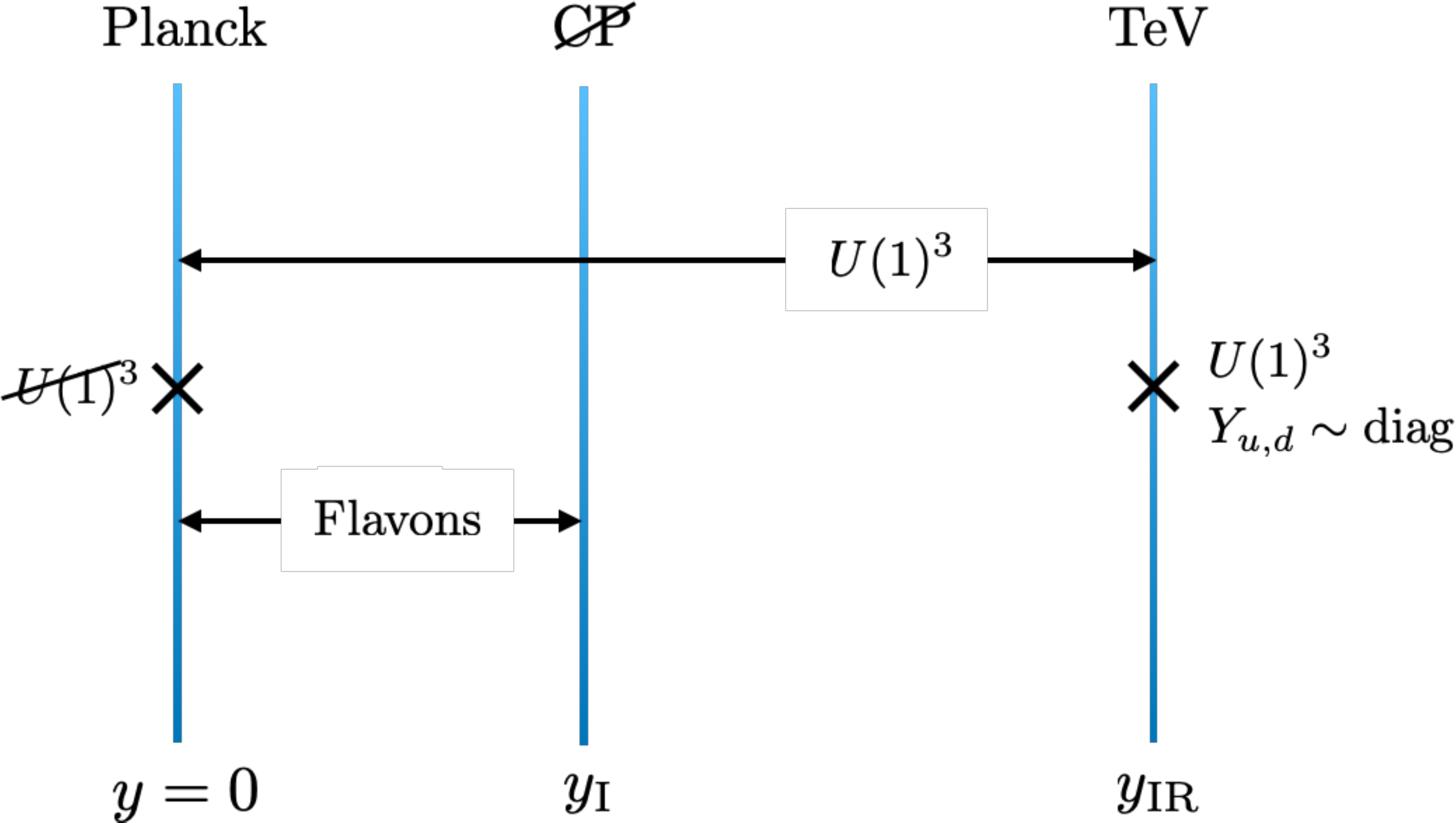}
   \vspace{0.2cm}
   \caption{A schematic picture of our setup for the $U(1)^3$ flavor symmetries.
   The $U(1)^3$ are spontaneously broken by VEVs of flavons in the subregion 1.
   The quark Yukawa couplings on the IR (TeV) brane are given by diagonal matrices
   due to the unbroken $U(1)^3$ symmetries.
   At the intermediate brane, the flavons couple to $\bar u$ and $\bar d$, which lead to the terms
   in Eqs.~\eqref{eq:mix_qd}, \eqref{eq:mix_qu}.
   The $U(1)^3$ symmetries are explicitly broken on the UV brane.
   }
\label{fig:flavor}
\end{figure}

Compared to the original BBP model,  our model contains KK excitation modes
corresponding to $q_{u,d}$, $\bar d$, $\bar u$, and the SM gauge bosons.
These KK excitation modes lead to new diagrams contributing to $\bar\theta$
together with flavor violation effects  as we will discuss in detail in the next section.
To avoid potentially large corrections to $\bar\theta$ and make our analysis simple, 
we introduce a flavor protection mechanism in the basis
shown in Eqs.~\eqref{eq:mu_0}, \eqref{eq:mix_qd}.\footnote{
The kinetic terms are canonically normalized. We call this basis as the gauge basis.}
We assume $U(1)^3=U(1)_1\times U(1)_2\times U(1)_3$ gauge symmetries in the whole bulk region
where each $U(1)$ rotates quarks with the same flavor index $f \, (=1,2,3)$.
Under the $U(1)_f$,
$\bar d_f$ has charges $(n\delta_{f1},n\delta_{f2},n\delta_{f3})$ with a positive integer $n$, $Q_f$ has charges $(-n\delta_{f1},-n\delta_{f2},-n\delta_{f3})$, and $\bar u_f$ has charges $(n\delta_{f1},n\delta_{f2},n\delta_{f3})$.
The $U(1)^3$ symmetries are spontaneously broken in the subregion 1 by VEVs of three flavons $\phi_f$
living only in the subregion $1$.
Here, $\phi_{f}$ has charges $(\delta_{1f},\delta_{2f},\delta_{3f})$ under the $U(1)_f$.
The $U(1)^3$ symmetries are also explicitly broken on the UV brane.\footnote{
If the explicit breaking terms are suppressed, we have axions~\cite{Ema:2016ops,Ema:2018abj}.}
Since the $U(1)^3$ symmetries are unbroken in the subregion 2,
the quark Yukawa couplings are given by diagonal matrices.%
\footnote{Note that the Yukawa couplings after integrating out the heavy vector-like quarks are not diagonal, and the CKM structure is obtained from the diaognal Yukawa terms and the $q-d$ $(q-u)$ mixing terms on the intermediate brane.}
The intermediate brane-localized terms in 
Eqs.~\eqref{eq:mix_qd}, \eqref{eq:mix_qu}
are suppressed by the flavon VEVs.
By taking $e.g.$ $n=5$ and $\langle \phi_f\rangle/M_5|_{y=y_{\rm I}}\approx 10^{-1}$, we obtain $a^{u,d}_{af}\approx10^{-5}$ which is enough to suppress corrections to $\bar\theta$ as we will discuss in the next section.
Fig.~\ref{fig:flavor} shows a schematic picture of our setup for the flavor symmetries.
At the intermediate brane, we can write down flavor off-diagonal kinetic terms of the $SU(2)_L$ singlet quarks
with some powers of the flavon VEVs. However, these terms are suppressed well for $e.g.$ $n=5$ and
$\langle \phi_f\rangle/M_5|_{y=y_{\rm I}}\approx 10^{-1}$,
$i.e.$ the ratio of the diagonal and off-diagonal kinetic terms is given by $\approx 10^{-10}$.

The quark mass hierarchies and the CKM structure are given from the diagonal Yukawa terms on the IR brane, and the terms on the intermediate brane of Eqs.~\eqref{eq:mix_qd} and \eqref{eq:mix_qu} in which the flavor symmetries are broken. As a demonstration, 
we assume the following couplings,
\begin{align}
&(Y^d_{11},~Y^d_{22},~Y^d_{33})\approx (1.8\times 10^{-4}, ~9.4\times 10^{-4},~4.3\times 10^{-2})\ ,\nonumber\\[1ex]
&(Y^u_{11},~Y^u_{22},~Y^u_{33})\approx (2.0\times 10^{-4}, ~2.6\times 10^{-2},~1)\ ,\nonumber\\[1ex]
&(a^d_{a1}\langle \eta_a\rangle ,~a^d_{a2}\langle \eta_a\rangle,~a^d_{a3}\langle \eta_a\rangle)\nonumber\\
&\approx (0.19+i 0.092 ,~-0.091+i 0.18,~0.65-i 0.41)m_{\rm I}\ ,\nonumber\\[1ex]
&(a^u_{a1}\langle \eta_a\rangle ,~a^u_{a2}\langle \eta_a\rangle,~a^u_{a3}\langle \eta_a\rangle)\nonumber\\
&\approx (-0.27-i 0.0084 ,~0.81+i 0.39,~0) m_{\rm I}\ \nonumber,\\[1ex]
&\mu_d \approx 0.034 m_{\rm I}\ ,~\mu_u\approx 0.0093 m_{\rm I}\ ,
\label{parameter_assumption}
\end{align}
where $Y^u_{ff'}$ denote the up-type quark Yukawa couplings,
the off-diagonal elements of $Y^{d,u}$ are zero due to the flavor symmetries,
and $m_{\rm I}$ is a mass parameter given by
$m_{\rm I}\approx  \langle\eta_a\rangle (\langle \phi_f\rangle/M_5)^n|_{y=y_{\rm I}}$.
These couplings are reasonably reproduced by taking $e.g.$ $k_1\approx k_2\approx k$, $k y_{\rm IR}=33$, $k y_{\rm I}=17$, $n=5$, $\langle \phi_f \rangle/M_5|_{y=y_{\rm I}}\approx 10^{-1}$,
and the following dimensionless bulk mass parameters of $\psi_{\bar d_f}, \psi_{\bar u_f}$ and $\psi_{q_{u,d}}$ in units of $k$,
\begin{align}
c_d\approx \left\{1,0.93,0.66\right\} , \quad 
c_u\approx\left\{1,0.7,0\right\} , \quad 
c_{q_{u,d}}\approx-1.2\ .
\end{align}
With the assumption of Eq.~\eqref{parameter_assumption},
the up and down-type quark masses and the CKM matrices are, respectively, given as
\begin{align}
m_u&\approx (1\,{\rm MeV},~1.3\,{\rm GeV},~173\,{\rm GeV})\ ,\nonumber\\[1ex]
m_d&\approx (5\,{\rm MeV},~120\,{\rm MeV},~2.7\,{\rm GeV})\ ,\nonumber\\[1ex]
|V_{\rm CKM}|&=
\left(
\begin{array}{ccc}
|V_{ud}| & |V_{us}| & |V_{ub}|\\
|V_{cd}| & |V_{cs}| & |V_{cb}|\\
|V_{td}| & |V_{ts}| & |V_{tb}|
\end{array}
\right) \\[1ex]
&\approx 
\left(
\begin{array}{ccc}
0.98 & 0.18 & 0.0074\\
0.18 & 0.98 & 0.039\\
0.015 & 0.037 & 0.99
\end{array}
\right)\ \nonumber .
\end{align}
We also get the Jarlskog invariant, $J\approx 1.2\times 10^{-5}$.

\section{Correction to $\bar\theta$}\label{theta_correction}
Let us first discuss how our warped extra dimension model can suppress contributions to $\bar\theta$ which are problematic in the non-supersymmetric BBP model. The model forbids the higher dimensional operators presented as in Eq.~\eqref{eq:danger_2} at the tree level
because $\bar q_{u,d}$ and $\eta_a$ live on different branes.
The first term in Eq.~\eqref{eq:danger_3} is also forbidden at the tree level by the same reason.
The quartic interactions of $\eta_a$ in Eq.~\eqref{eq:danger_3}, however, cannot be erased in the similar way
because they live on the same brane. 
The contribution to $\bar\theta$ from those interactions was estimated in Eq.~\eqref{eq:delta_theta_gamma},
and a simple solution to satisfy the condition $\delta\bar\theta\leq 10^{-10}$ is
to assume the coefficients $a^{u,d}_{af}$ are much smaller than
$\mathcal{O}(1)$ like $a^{u,d}_{af}\lesssim 10^{-3}$. 
Such small $a^{u,d}_{af}$ are realized by the flavor symmetries as discussed in the previous section.

We now explore radiative corrections to $\bar\theta$ that are specific
to our 5D multi-brane model.
It is important to note that all CP phases given from Eq.~\eqref{eq:mix_qd} can be moved into the up-quark sector of Eq.~\eqref{eq:mix_qu} completely (or vise versa) by a phase rotation,
$Q_f\to e^{i\alpha_f}Q_f$, $\bar d_f\to e^{-i\alpha_f}\bar d_f$ and $\bar u_f\to e^{-i\alpha_f}\bar u_f$
where $\alpha_f$ are determined to erase the phases in $a^d_{af}\langle\eta_a\rangle$. 
Note that only a pair of vector-like quarks is introduced in each of the down- and up-type sectors,
and the parameters $\mu_{u,d}$ are $1\times 1$ matrices, which make it possible
to move all the CP phases in the down-sector into the up-type sector.
This implies that one needs both up and down-type quarks in a diagram to obtain a non-zero correction to $\bar\theta$.
We define corrections to the tree-level up and down-type quark mass matrices $M_{u,d}$
as $\delta M_{u,d}$ so that
\begin{align}
\bar\theta&=\arg\det(M_u+\delta M_u)\det(M_d+\delta M_d) \nonumber \\
\label{eq:approx_theta_bar}
\begin{split}
&={\rm Im}  {\rm Tr}(R_u^\dagger \delta M_u L_u \mu_u^{-1}+R_d^\dagger \delta M_d L_d \mu_d^{-1})  \\
&\quad +\mathcal{O}(\delta M_u^2,\delta M_d^2,\delta M_u\delta M_d)\ ,
\end{split}
\end{align}
where $R_{u,d}, L_{u,d}$ denote unitary matrices diagonalizing $M_{u,d}$, $i.e.$ $R_u^\dagger M_u L_u=\mathcal{M}_u$ and $R_d^\dagger M_d L_d=\mathcal{M}_d$ 
where  $\mathcal{M}_u$ and $\mathcal{M}_d$ are real and diagonal matrices.
Note that $M_{u,d}$ here are different from $\hat M_{u,d}$ introduced around Eq.~\eqref{eq:M_d}, $i.e.$ $M_{u,d}$ are the up-type and down-type quark matrices including $q_{u,d}$, $\bar q_{u,d}$, the SM up and down-type quarks
and their KK modes, respectively.
The last line of Eq.~\eqref{eq:approx_theta_bar} denotes terms
with higher orders of $M_{u,d}$.
Since we have not found important corrections to $\bar \theta$ through these terms,
we focus on corrections through the leading order terms in the following discussion.

At the one-loop level, there is no diagram containing both up and down-type quarks
together with SM gauge bosons or their KK excitation modes.
A correction to $\delta M_d$ may come from diagrams involving $h^\pm$ (NG bosons),
which are however leading to \cite{Cheung:2007bu}
\begin{align}
\label{eq:theta_one_loop}
\bar \theta& \propto
\sum_{n=0}^\infty {\rm Im}  {\rm Tr}\left(R_d^\dagger Y_d M_u^\dagger (M_u M_u^\dagger)^n Y_uL_d\mathcal{M}_d^{-1}\right) \nonumber \\
&\propto\sum_{n=0}^\infty {\rm Im}\sum_i (\mathcal{M}_d L_d^\dagger P_0 L_d \mathcal{M}_d^{2n+2}L_d^\dagger P_0 L_d)_{ii}/(\mathcal{M}_d)_{ii} \nonumber \\
&=0\ .
\end{align}
Here, $Y_d$ is the SM down-type Yukawa matrix and $P_0$ denotes a matrix projecting $\mathcal{M}_d$ on $v Y_d$, $i.e.$ $vY_d\equiv R_d\mathcal{M}_dL_d^\dagger P_0$.
In the last equality of Eq.~\eqref{eq:theta_one_loop},
we have used the fact that $V_{ij}V^\dagger_{ji}$ is real for a matrix $V$.%
\footnote{In other words, a one-loop diagram cannot contain both terms of Eqs.~\eqref{eq:mix_qd}, \eqref{eq:mix_qu}.}
A one-loop correction to $\delta M_u$ involving $h^\pm$ also vanishes by the same reason.
Therefore, we conclude that there is no sizable correction to $\bar\theta$ at the one-loop level.

\begin{figure}
\centering
\hspace{-0.7cm}
   \includegraphics[width=0.9\linewidth]{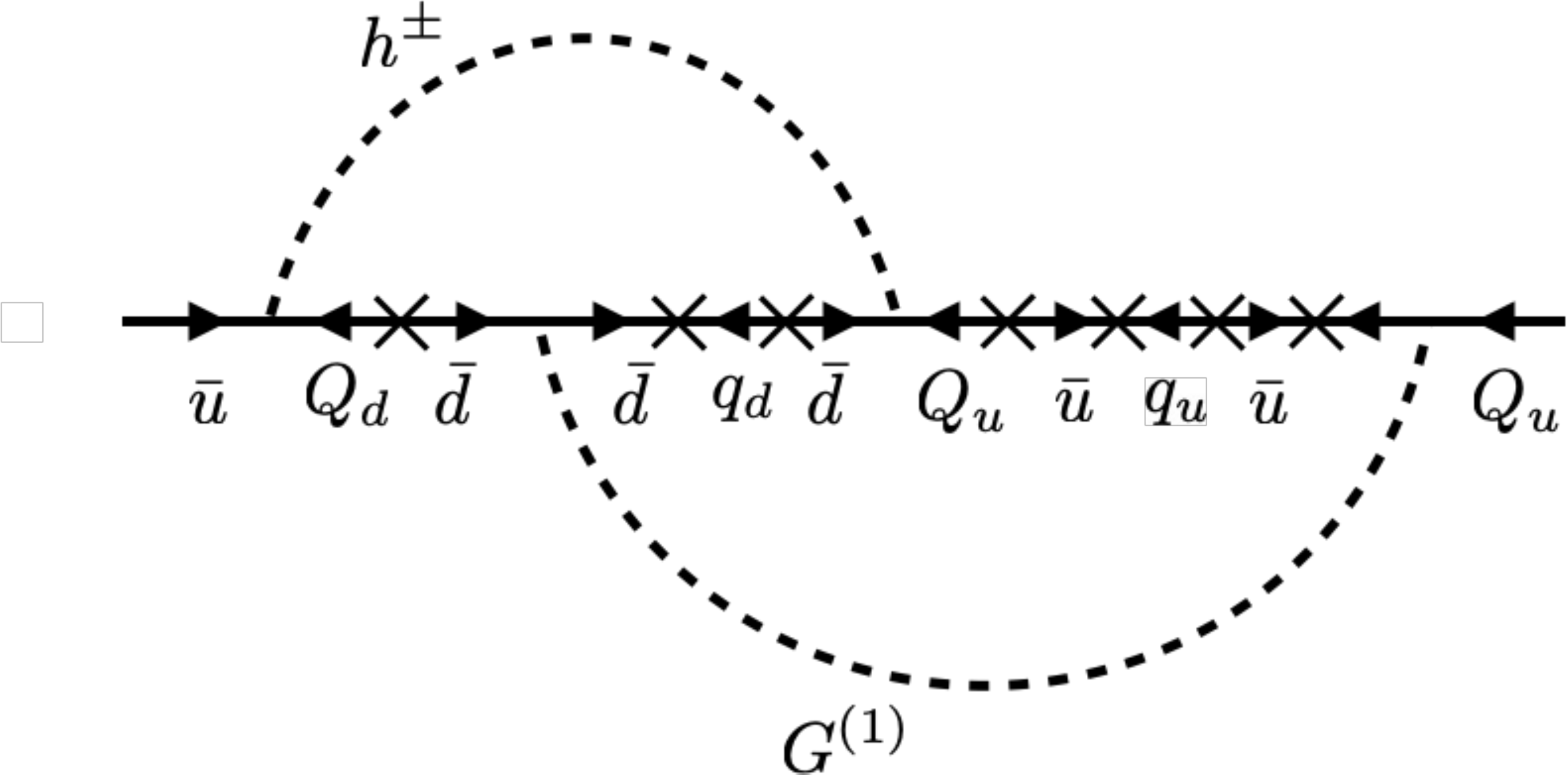}
   \vspace{0.2cm}
   \caption{A two-loop diagram involving $h^\pm$ and the first KK excitation of the gluon $G^{(1)}$.
   Here, $\bar d$, $\bar u$, $q_d$ and $q_u$ include KK excitation modes as well as zero modes.
   $Q_{u,d}$, respectively, denote the up and down-type quark components of $Q$.
   We have omitted the flavor indices for notational simplicity.
   The solid lines represent fermion lines, and the dashed lines denote boson lines including gauge bosons and Higgs.}
\label{fig:hpm_G_1}
\end{figure}

\begin{figure}
\centering
\hspace{1cm}
   \includegraphics[width=0.9\linewidth]{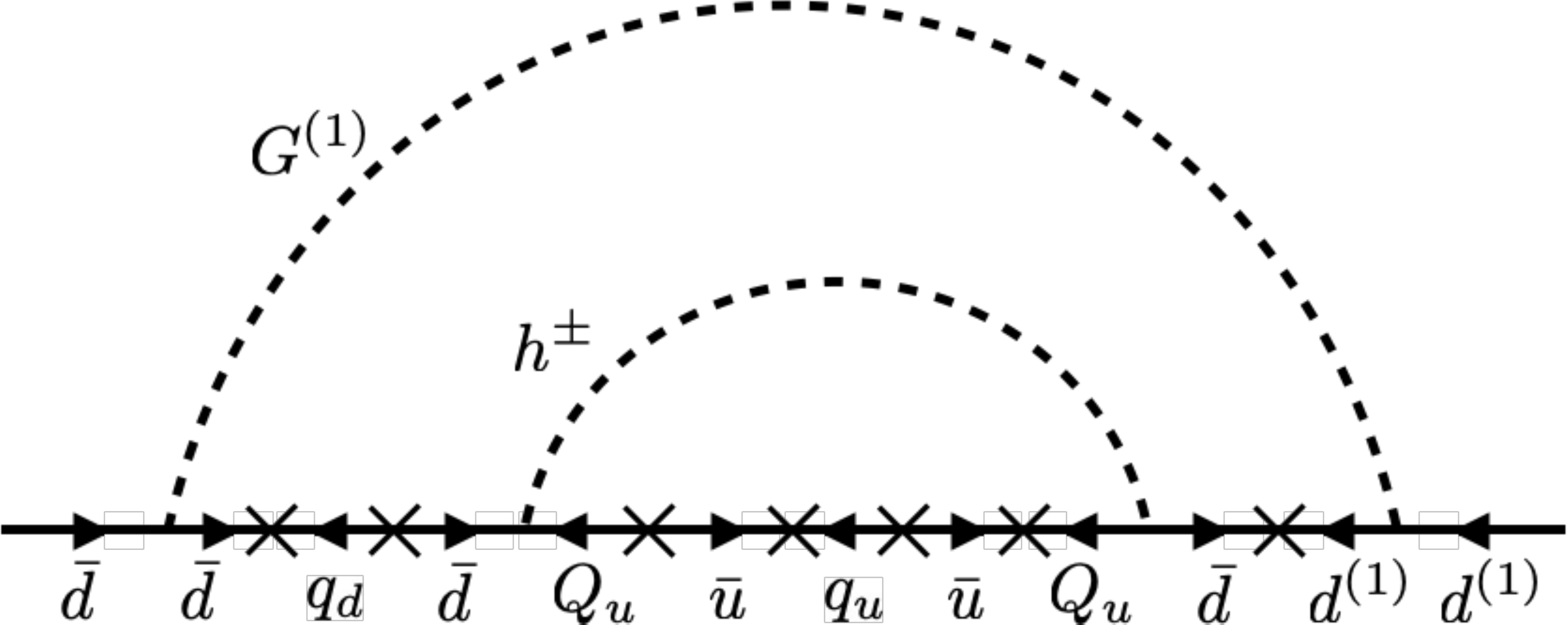}
   \vspace{0.2cm}
   \caption{A two-loop diagram which contains the first KK excitation mode of the down-type quark $d^{(1)}$
   (the Dirac partner of $\bar d^{(1)}$).
   The flavor indices are also omitted.
   }
\label{fig:hpm_G_2}
\end{figure}

At the two-loop level, the number of diagrams which may contribute to $\bar\theta$ drastically increases.
To find relevant diagrams involving both up and down-type quarks,
we require at least one $h^\pm$ or $W^\pm$ in a diagram.
In our warped extra dimension model, KK excitation modes of the SM gauge bosons, up and down-type quarks
and $q_{u,d}$ can enter into a diagram.
Fig.~\ref{fig:hpm_G_1} shows a diagram which may lead to an important contribution to $\bar\theta$.
The diagram is given in the gauge basis and contains $h^\pm$ and the first KK excitation of the gluon $G^{(1)}$ in the loops.
A correction to $\bar\theta$ generated by this diagram is however suppressed by two SM Yukawa couplings
and three SM quark mass insertions.
Moreover, the bulk profiles of the down-type quark zero modes to explain the mass hierarchies give
a suppression for their couplings to the KK gluon whose profile is localized toward the IR brane.
Therefore, this diagram safely leads to $\bar\theta<10^{-10}$.
Fig.~\ref{fig:hpm_G_2} is a diagram with different topology
which contains the first KK excitation mode of the down-type quark (the Dirac partner of $\bar d^{(1)}$).
This diagram also includes $\eta \bar d_f^{(1)} q_d$, $\eta \bar u_3^{(0)} q_u$ and $\eta \bar u_f^{(1)} q_u$ couplings 
suppressed by their profiles.
To obtain a sufficient suppression, we require $\exp(-k y_{\rm IR})/\exp(-k y_{\rm I})\ll 10^{-5}$
where we assume $k_1\approx k_2\equiv k$.
Consequently, we require two conditions to realize $\bar\theta<10^{-10}$ at the two-loop level:
the mass hierarchies of SM quarks are explained by their bulk profiles,
and a distance between the intermediate and IR branes is large enough, $e^{k (y_{\rm I}-y_{\rm IR})}\ll 10^{-5}$.
Under these conditions, we have also investigated diagrams involving
$e.g.$ $h^\pm-h^\pm$, $h^\pm-h^0$, $h^\pm-W^\pm$, $W^\pm-h^0$
and found no dangerous corrections to $\bar\theta$.

\begin{figure}
\centering
\hspace{0cm}
   \includegraphics[width=0.55\linewidth]{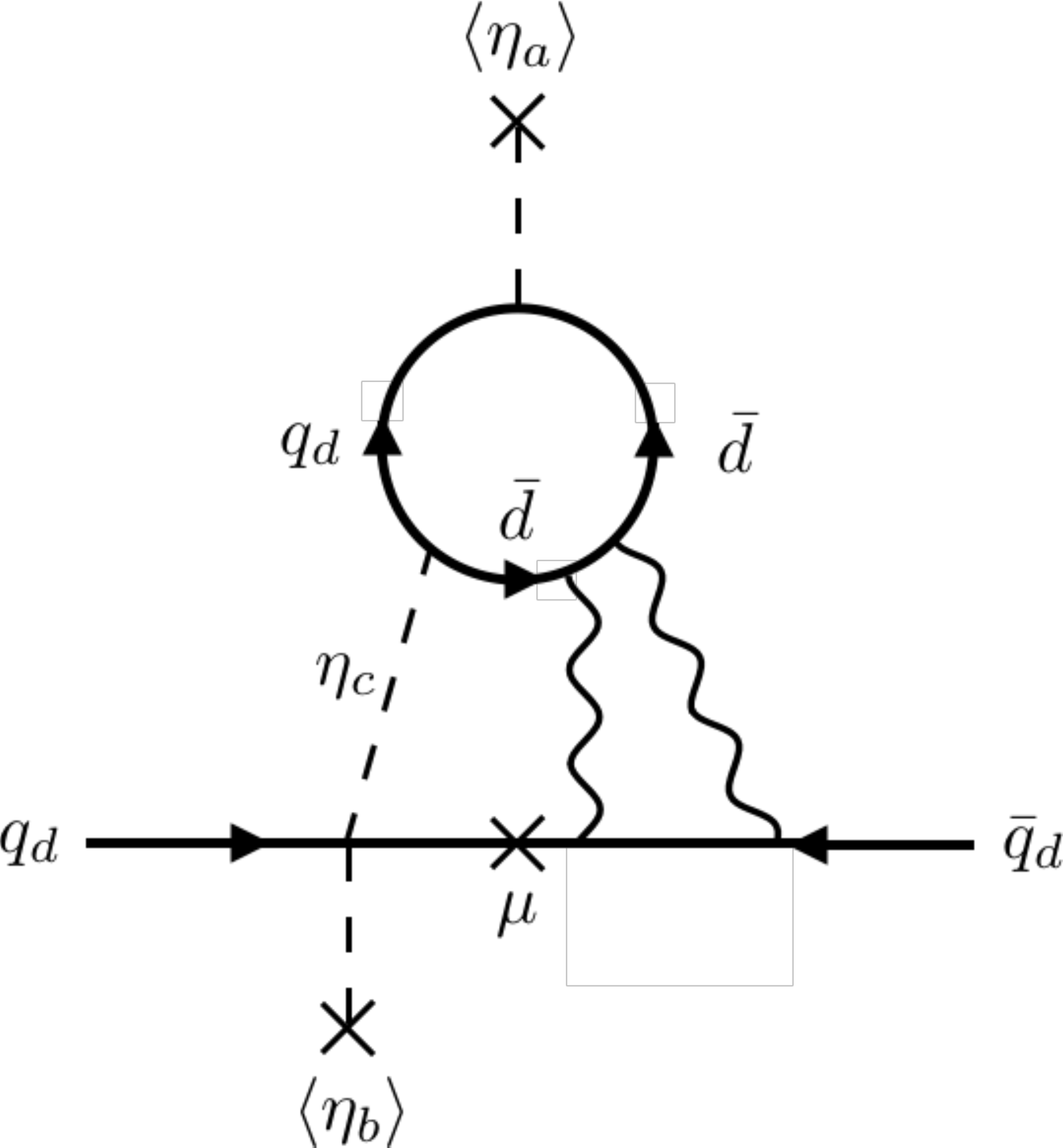}
   \caption{An example diagram contributing to $\bar\theta$.
   The wavy lines denote the SM gauge bosons.}
\label{fig:diagram}
\end{figure}

The kinetic term of $q_{u,d}$ can be modified by interactions with $\eta_a$ on the intermediate brane,
\begin{align}
\mathcal{L} \supset i Z_{u,d}\left(\eta_a/M_{\rm I},\eta_a^*/M_{\rm I}\right) q_{u,d}^\dagger\bar\sigma^\mu\mathcal{D}_\mu q_{u,d}\ ,
\end{align}
where $M_{\rm I}\approx k_1 e^{-k_1y_{\rm I}}$ represents a cutoff scale of the intermediate brane.
Since $Z_{u,d}\left(\langle\eta_a\rangle/M_{\rm I},\langle\eta_a^*\rangle/M_{\rm I}\right)$ is a real number,
it does not provide a new correction to $\bar\theta$.
However, $Z_{u,d}\left(\eta_a/M_{\rm I},\eta_a^*/M_{\rm I}\right)$ may contribute to $\bar\theta$
through diagrams such as the one shown in Fig.~\ref{fig:diagram}.
This diagram leads to a correction to the $\mu q_d\bar q_d$ term,
$i.e.$ a CP phase is provided in the diagonal component of $M_d$.
The correction to $\bar\theta$ is estimated as
\begin{align}
|\delta \bar \theta|\sim \left|\sum_{a,b,c,f}\left(\frac{1}{16\pi^2}\right)^3 \frac{ \langle\eta_a\rangle  \langle\eta_b\rangle a_{af}^d a_{cf}^d}{\Lambda_{\rm I}^2}  \right|\ ,
\end{align}
where
we have assumed coefficients in $Z_d\left(\eta_a/M_{\rm I},\eta_a^*/M_{\rm I}\right)$ are $\mathcal{O}(1)$. 
This correction can be sufficiently suppressed by $a^d_{af}\lesssim 10^{-3}$ that is required to suppress the contribution from Eq.~\eqref{eq:delta_theta_gamma}.
We note that modified kinetic terms of $\bar u$ and $\bar d$ do not give relevant corrections
because their off-diagonal components are suppressed by the flavor symmetries introduced in Sec.~\ref{model}.

Let us comment on corrections from KK modes of the $U(1)^3$ flavor gauge fields and flavons.
The former corrections are given by replacing the KK gluon in Fig.~\ref{fig:hpm_G_1}
with a KK mode of the $U(1)^3$ gauge field.
The corrections depend on the $U(1)^3$ gauge couplings,
and if the couplings are tiny, they are suppressed. 
The flavons $\phi$ couple to $\bar d$, $\bar u$ in Eqs.~\eqref{eq:mix_qd} , \eqref{eq:mix_qu}
like $\mathcal{L}\sim \phi^{*5} \eta \psi_{\bar d} \psi_{q_d}+\phi^{*5} \eta \psi_{\bar u} \psi_{q_u}+\rm h.c.$,
as well as the $\bar d$, $\bar u$ kinetic terms.
The corrections are suppressed for small flavon VEVs, $\langle \phi \rangle/M_5|_{y=y_{\rm I}}\ll 1$,
and large flavor charges of quarks, $n\gg 1$.
Therefore, we conclude that the corrections from KK modes of the flavor gauge bosons and flavons can be sufficiently small and hence do not pose a problem.

\section{Conclusions}\label{conclusion}

We have explored a doubly composite dynamics via a holographic warped extra dimension model with three 3-branes
that realizes the NB mechanism to address the strong CP problem.
The scale of spontaneous CP violation was introduced as the scale of the intermediate brane.
We have shown that three branes are essential to forbid dangerous operators leading to a large $\bar \theta$.
With the help of flavor symmetries for quarks, the model can solve the strong CP problem without fine-tuning.
The SM quark masses and mixings are naturally explained by the bulk profiles of the quarks.

%
\section*{Acknowledgements}

S.G. acknowledges support from the U.S. National Science
Foundation Grant NSF-PHY-1915093.
S.L. was supported by the National Research Foundation of Korea (NRF) grant funded by the Korea government (MEST)
(No. NRF-2021R1A2C1005615) and the Visiting Professorship at Korea Institute for Advanced Study.
S.L. is also partly supported by the Korea-CERN program.
Y.N. is supported by Natural Science Foundation of China under grant No.~12150610465.
M.S. thanks ICRR for their hospitality.

%

\appendix 
\section{5D fermion in three 3-brane models}
\label{bulkfermion}

We consider a Dirac fermion $\Psi$ living in the whole 5D spacetime whose metric is given in Eq. (8) in the main text.
The action is 
\begin{align}
S&=\int_0^{y_{\rm I}} dy\int d^4x \sqrt{g}\left[\frac{i}{2}\left(\bar\Psi\Gamma^M\partial_M\Psi-\partial_M\bar\Psi \Gamma^M\Psi\right)\right.\nonumber\\
&\left.~~~-m_{\Psi,1}\bar\Psi\Psi\right] \nonumber \\[1ex]
&+\int_{y_{\rm I}}^{y_{\rm IR}} dy\int d^4x \sqrt{g}\left[\frac{i}{2}\left(\bar\Psi\Gamma^M\partial_M\Psi-\partial_M\bar\Psi \Gamma^M\Psi\right)\right.\nonumber\\
&\left.~~~-m_{\Psi,2}\bar\Psi\Psi\right] ,
\label{fermionaction}
\end{align}
where $M = (\mu, y)$ and we have defined
\begin{align}
\Gamma^M\equiv e^M_{~A}\gamma^A\ , \qquad g^{MN}\equiv e^M_{~A}e^N_{~B}\eta^{AB}\ ,
\end{align}
with the vielbein $e^M_{~A}$ which is non-zero for diagonal components, $i.e.$ $e^\mu_{~\alpha}=e^{A(y)}\delta^{\mu}_\alpha,~e^y_{~5}=1$.
The gamma matrices are 
\begin{align}
&\gamma^A \equiv(\gamma^\mu,~-i\gamma_5) ,~
\gamma^\mu\equiv
\left(
\begin{array}{cc}
0 & \sigma^\mu\\
\bar\sigma^\mu & 0
\end{array}
\right) ,~
\gamma_5\equiv\left(
\begin{array}{cc}
-{\bf 1} & 0\\
0 & {\bf 1}
\end{array}
\right) ,\\
&\sigma^\mu\equiv(\bm{1},~\bm{\sigma}) \, , \qquad \bar \sigma^\mu\equiv(\bm{1},-\bm{\sigma}) \, .
\end{align}
Here, ${\bm {\sigma}}$ denote the Pauli matrices.
Note that the dependence on the spin connection  is canceled in the equation of motion
as in the case of the two 3-brane setup.

The bulk equation of motion is obtained from the variation of the action \eqref{fermionaction},
\begin{align}
&e^{k_{1,2} y}i\gamma^\mu \partial_\mu\Psi_--\partial_5 \Psi_++k_{1,2}(-c_{1,2}+2)\Psi_+=0\ ,\\
&e^{k_{1,2} y} i\gamma^{\mu}\partial_\mu\Psi_++\partial_5 \Psi_--k_{1,2}(c_{1,2}+2)\Psi_-=0\ ,
\end{align}
where $m_{\Psi_{1,2}} \equiv c_{1,2}k_{1,2}$ and $\Psi_{\pm}$ denote the left and right-handed spinors defined as
\begin{align}
&\Psi=\left(
\begin{array}{c}
\psi_+\\
\psi_-
\end{array}
\right) ,
~\Psi_+=\left(
\begin{array}{c}
\psi_+\\
0
\end{array}
\right) ,
~\Psi_-=\left(
\begin{array}{c}
0\\
\psi_-
\end{array}
\right) , \\[1ex]
&\Psi_\pm=\mp\gamma_5\Psi_\pm\ .
\end{align}
By using the Dirac equation $i\gamma^\mu\partial_\mu\Psi_{\pm}^{(n)}=m_n\Psi_{\pm}^{(n)}$, 
the above equations are further reduced to
\begin{align}
&\partial_5 \psi_+^{(n)}+k_{1,2}(c_{1,2}-2)\psi_+^{(n)}=e^{k_{1,2} y}m_n \psi_-^{(n)}\ ,\\[1ex]
&-\partial_5 \psi_-^{(n)}+k_{1,2}(c_{1,2}+2)\psi_-^{(n)}=e^{k_{1,2} y}m_n \psi_+^{(n)}\ .
\end{align}
In the variation of the action, the boundary terms on the UV and IR branes vanish when
\begin{align}
\delta\bar\Psi_+\,\Psi_-=0 \quad \text{and} \quad \delta\bar\Psi_-\,\Psi_+=0 
\end{align}
are satisfied while the boundary term on the intermediate brane vanishes for
\begin{align}
\delta\bar\Psi_+\,[\Psi_-]|_{y=y_{\rm I}}=0 \quad \text{and} \quad \delta\bar\Psi_-\,[\Psi_+]|_{y=y_{\rm I}}=0\ ,
\end{align}
where $[X]|_{y=y_{\rm I}} \equiv \lim_{\epsilon\to0}X(y_{\rm I}+\epsilon)-X(y_{\rm I}-\epsilon)$.
Thus, we require
\begin{align}
&\psi_+|_{y=y_{\rm UV,IR}}=0 \quad \text{or} \quad \psi_-|_{y=y_{\rm UV,IR}}=0 \ , \\[1ex]
&[\psi_+]|_{y=y_{\rm I}}=0 \quad \text{and} \quad [\psi_-]|_{y=y_{\rm I}}=0 \ ,
\end{align}
as boundary conditions at the three branes.
By taking the Dirichlet condition on the UV and IR branes for $\Psi_+$ $(\Psi_-)$, 
the zero mode solution for $\Psi_+$ $(\Psi_-)$ becomes trivial, 
$i.e.$ $\Psi_+=0$ $(\Psi_-=0)$ for the whole space.
Then, only the right-handed (left-handed) spinor appears as a massless mode as given in the two 3-brane case.

\bibliography{bib}
\bibliographystyle{utphys}

\end{document}